\documentclass[prl,aps,nopacs,superscriptaddress,nofootinbib]{revtex4}
\usepackage{graphicx} % Include figure files
\usepackage{dcolumn}  % Align table columns on decimal point
\usepackage{bm}       % bold math
\usepackage{amsmath}
\usepackage{epsfig}
\begin{document}

\title{Magneto-optical response in bimetallic metamaterials}
\author{Evangelos~Atmatzakis}
\affiliation{Optoelectronics Research Centre and Centre for Photonic Metamaterials, University of Southampton, Southampton SO17 1BJ, United Kingdom}
\author{Nikitas~Papasimakis}
\affiliation{Optoelectronics Research Centre and Centre for Photonic Metamaterials, University of Southampton, Southampton SO17 1BJ, United Kingdom}
\author{Vassili~Fedotov}
\affiliation{Optoelectronics Research Centre and Centre for Photonic Metamaterials, University of Southampton, Southampton SO17 1BJ, United Kingdom}
\author{Guillaume~Vienne}
%\affiliation{Data Storage Institute, Agency for Science Technology and Research (A*STAR), Singapore 117608}
\affiliation{School of Electrical and Electronic Engineering, Nanyang Technological University, Singapore 639798}
\author{Nikolay~I.~Zheludev}
\email{niz@orc.soton.ac.uk}
\affiliation{Optoelectronics Research Centre and Centre for Photonic Metamaterials, University of Southampton, Southampton SO17 1BJ, United Kingdom}
\affiliation{The Photonics Institute and Centre for Disruptive Photonic Technologies, Nanyang Technological University, Singapore 637371}
\date{\today}

\begin{abstract}
 We demonstrate resonant Faraday polarization rotation in plasmonic arrays of bimetallic nano-ring resonators consisting of Au and Ni sections. This metamaterial design allows to optimize the trade-off between the enhancement of magneto-optical effects and plasmonic dissipation. Although Ni sections correspond to as little as $\sim6\%$ of the total surface of the metamaterial, the resulting magneto-optically induced polarization rotation is equal to that of a continuous film. Such bimetallic metamaterials can be used in compact magnetic sensors, active plasmonic components and integrated photonic circuits.  
\end{abstract}

\maketitle

% ---------- Introduction ----------
The ability to tailor light-matter interactions is equally important for the development of current and future technologies (telecommunications, sensing, data storage), as well as for the study of the fundamental properties of matter (spectroscopy). A typical example involves the exploitation of  magneto-optical (MO) effects, where quasistatic magnetic fields can induce optical anisotropy in a material. This is a direct manifestation of the Zeeman effect, the splitting of electronic energy levels due to interactions between magnetic fields and the magnetic dipole moment associated with the orbital and spin angular momentum \cite{Zve_Kot}. This energy splitting gives rise to numerous polarization phenomena, such as magnetically-induced birefringence and dichroism, which enable dynamic control over the polarization state of light. 
 
In recent years, magnetoplasmonics, the study of systems that combine plasmonic and magnetic properties of matter, holds promise to both enhance MO effects and enable non-reciprocal, magnetic-field-controlled plasmonic devices ~\cite{PhysRevLett.73.3584,gonzalez-diaz_plasmonic_2008,:/content/aip/journal/apl/97/26/10.1063/1.3533260,doi:10.1021/nl1042243,Chin2013,du_shape-enhanced_2010}. Ferromagnetic metals, such as Fe, Ni, and Co, are known to be magneto-optically active but have poor plasmonic properties in the near-infrared (NIR) spectral range due to Joule losses. In contrast, noble metals suffer less from losses but exhibit negligible MO response. By combining ferromagnetic and noble metals, one can construct a system with strong MO response. Indeed, early studies identified dramatic enhancement of the MO response due to the strong localized electromagnetic fields associated with plasmonic resonances. In particular, enhanced Faraday effect (FE) and magneto-optical Kerr effect (MOKE) have been observed in hybrid devices that simultaneously support plasmonic resonances and are MO active ~\cite{feil_magneto-optical_1987,Reim1988,Ferre1990}. Furthermore, nanostructuring of magnetoplasmonic systems provides control over the frequency dispersion and phase of MO effects ~\cite{Chin2013,gonzalez-diaz_plasmonic_2008,du_shape-enhanced_2010,Belotelov2011,maccaferri_ultrasensitive_2015}, while maintaining a compact form factor. In bulk media MO effects depend on the length of interaction between light and matter. However, the light confinement originating from nanostructuring can offer similar response at significantly smaller dimensions, an often required property for the realization of compact modulators, isolators, and circulators in telecommunications and photonic devices ~\cite{1291717}. On the other hand, the presence of external magnetic fields can be employed to control the plasmonic response, a desirable functionality in sensing applications ~\cite{Sepulveda:06,doi:10.1021/acs.nanolett.5b00372}. 

\begin{figure*}[htb!]
	\includegraphics[width=1\textwidth]{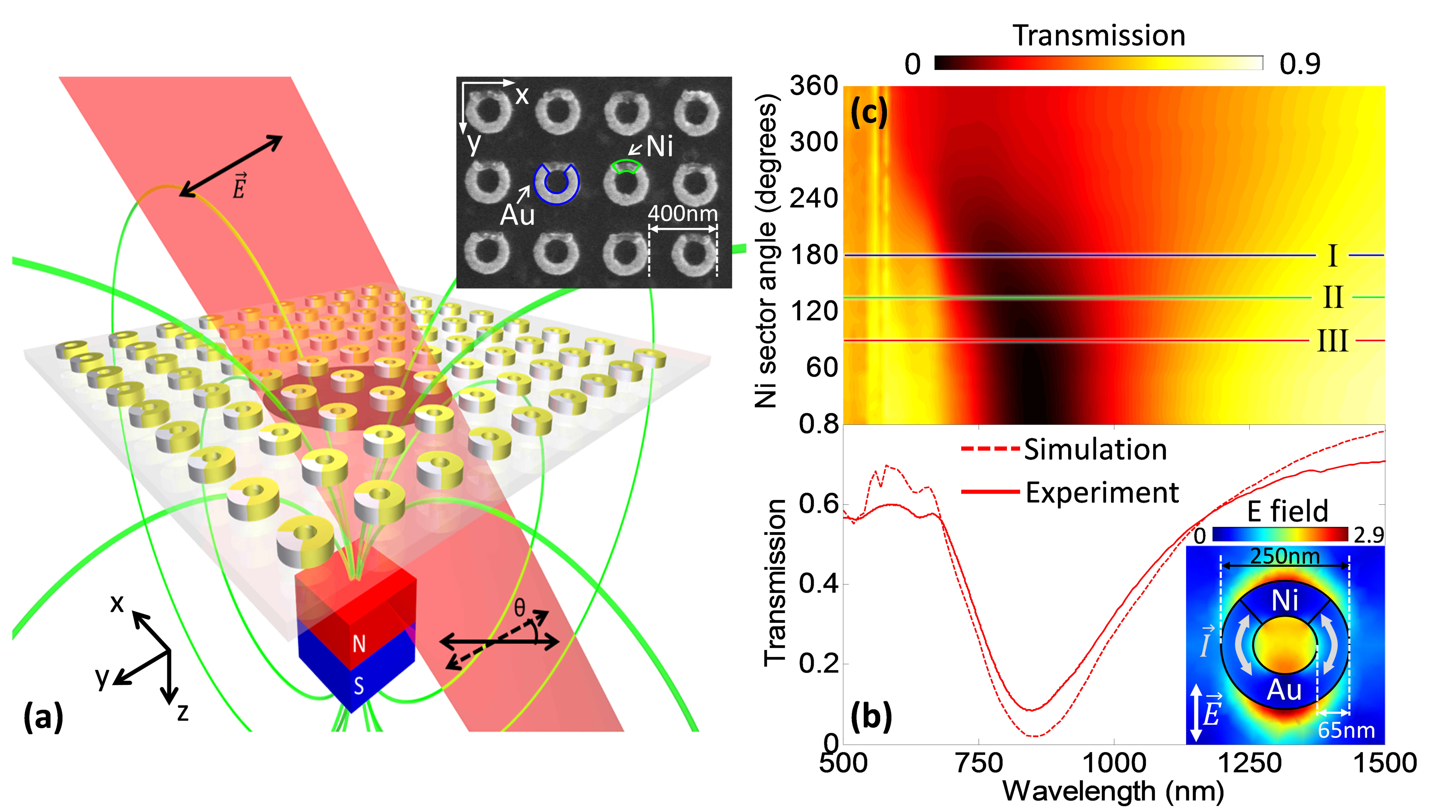}
	\caption{{\bf Linear optical properties of bimetallic ring resonator arrays.} {\bf (a)} Schematic of the bimetallic metamaterial array. In the presence of an external magnetic field the metamaterial induces rotation of the polarization azimuth angle, $\theta$, on the incident beam. The inset is a scanning electron microscope (SEM) image of a fabricated sample. {\bf (b)} Characteristic transmission spectra for a metamaterial sample with Ni sector that spans $90^{\circ}$. The dashed line is obtained by a computational analysis of an infinite two-dimensional array, while the solid line corresponds to the experimentally measured spectra of a $100 \times 100\ \mu$m$^2$ metamaterial sample. {\bf Inset:} Electric field distribution in the vicinity of the metamaterial at resonance ($\lambda=850\,$nm). {\bf (c)} Transmittance of regular arrays consisting of bimetallic resonators with Ni sector from $0^{\circ}$ to $360^{\circ}$, calculated numerically. The three coloured lines correspond to different fabricated samples ($90^{\circ}$ - red/III, $135^{\circ}$ - green/II and $180^{\circ}$ - blue/I).}
	\label{Fig2}
\end{figure*}

Magnetoplasmonic metamaterials, consisting of arrays of plasmonic resonators hybridized with MO materials, employ resonant plasmonic fields to maximize the enhancement of MO effects. The MO active material can be introduced either as substrate/superstrate or as part of the plasmonic resonator. Since the enhancement of MO effects occurs mainly where the resonant plasmonic fields overlap with the MO active components, very compact magnetoplasmonic metamaterials can be realized. Here, we implement for the first time a design for magnetoplasmonic metamaterials where the MO active component is integrated directly into the plasmonic resonators. The MO response is provided by small Ni sections, which are combined with a gold split-ring to form a bimetallic ring resonator (as illustrated in Fig.\ \ref{Fig2}b). This design allows to optimize the trade-off between  dissipation loss (which leads to weaker local plasmonic fields) and the strength of the MO response, with both effects increasing for larger Ni sections. 
Moreover, varying the wavelength or polarization of the incident wave allows one to shift the area of light confinement along the resonator and thus control the MO response. Finally, our approach is based on continuous metallic nanostructures and enables not only to realize complex resonators with prescribed MO and plasmonic response, but also to exploit other types of physical response involving thermal and electric effects \cite{thermo,Guillaumme}.

% ---------- Results and Discussion ----------
\section{Results and discussion}
The materials of choice for the bimetallic metamaterial system are Au and Ni which comprise the two arc sectors that form the complete ring (Fig.\ \ref{Fig2}a). The angle, over which the Ni sector spans, controls the composition of the ring. Three samples have been fabricated by electron-beam lithography (see Methods for details) with Ni sectors of $90^{\circ}$, $135^{\circ}$ and $180^{\circ}$. The mean diameter of the rings is $185\pm 5\,$nm with a linewidth of $65\,$nm and a height of $60\,$nm. Each sample comprises a regular array of $250 \times 250$ unit cells, positioned with a period of $400\,$nm and covering a total area of $100 \times 100\ \mu$m$^2$.

A strong plasmonic resonance can be excited in the metamaterial under illumination with light polarized in the plane of symmetry of the resonators (yz plane - see Fig.\ \ref{Fig2}a). The characteristic case of a metamaterial with a $90^{\circ}$ Ni sector is presented in Fig.\ \ref{Fig2}b, where the resonance manifests as a transmission dip at a wavelength of $\sim 850\,$nm. The experimental transmission spectrum (solid red line) is in good agreement with the results of numerical modelling (dashed red line). At resonance, the electric field profile exhibits two "hot spots" of charge concentration, while currents oscillate along the y-axis (see inset to Fig.\ \ref{Fig2}b). This configuration corresponds to the lowest-order electric dipole mode of the ring. Scattering and dissipation in the system is determined by the regions of high current density. Strong currents in the Ni sector reduce the total power scattered at resonance, since nickel is less conductive and exhibits higher losses than gold due to Joule heating. In accordance, increasing the Ni sector size leads to damping of the resonance and a decrease of its Q-factor as it can be deduced from Fig.\ \ref{Fig2}c. %(varies between $\sim 2.5$ and $1.5$). 
Such changes in composition also alter the effective permittivity of the hybrid structure and shift the position of the resonance towards shorter wavelengths.

\begin{figure}
	\includegraphics[width=0.5\textwidth]{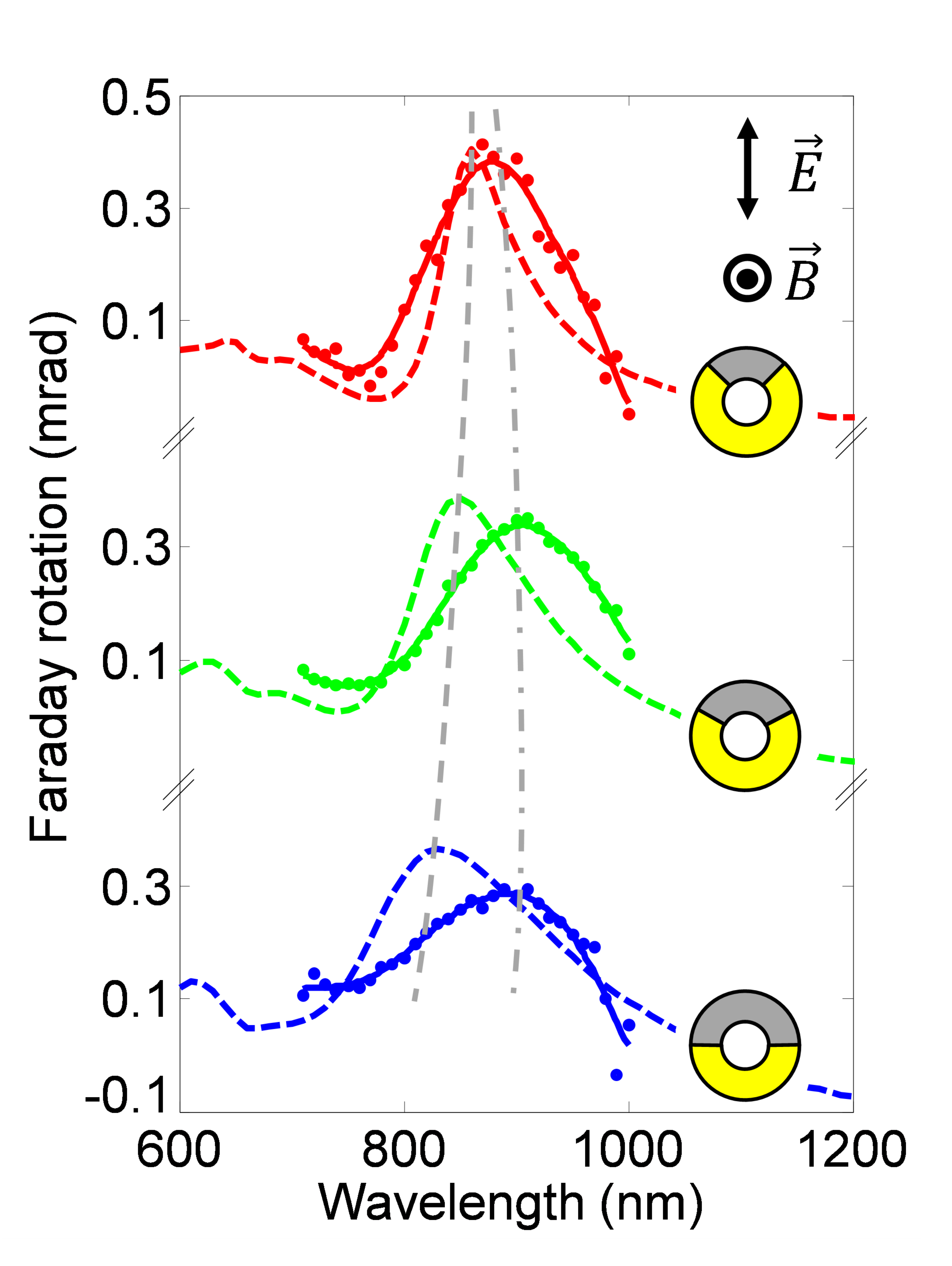}
	\caption{{\bf Faraday rotation of polarization azimuth.} Dots describe the experimental results on Faraday rotation of polarization azimuth for samples with Ni sectors that span over $90^{\circ}$ (red), $135^{\circ}$ (green) and $180^{\circ}$ (blue) in the presence of a $100\,$mT external magnetic field. Solid lines act as guide for the eye. Dashed lines represent the corresponding computational results when the linear optical response has been normalized to the experimental values (see methods). The dashed grey line follows the peak position of simulated data, as seen in Fig.\ \ref{Fig5}a, while dot-dashed line shows the corresponding trend for the experimental results.} 
	\label{Fig3}
\end{figure}

In presence of an external magnetic field (B) the permittivity tensor of Ni becomes non-diagonal, which affects the polarization state of light that is transmitted through the sample. The rotation of polarization azimuth, which is induced by the magnetic field, is experimentally measured using a highly sensitive polarimeter setup (discussed in Methods) and the results are shown in Fig.\ \ref{Fig3}. For all three lengths of the Ni sector ($90^{\circ}, 135^{\circ}$ and $180^{\circ}$), the Faraday rotation spectra follow closely the linear optical response with the maxima of rotation occurring close to the plasmonic resonance frequency. Importantly, smaller Ni sectors lead to a stronger Faraday effect: reducing by $50\%$ the arc length of the Ni sector (from $180^{\circ}$ to $90^{\circ}$) leads to a $40\%$ increase of peak rotation (from $0.29\,$mrad to $0.41\,$mrad). Simultaneously, the linewidth of the Faraday rotation resonance decreases significantly. These features are reproduced by finite element simulations (dashed lines) following a normalization of numerical transmittance spectra to the experimental ones (see Methods for details). In both experiment and theory, the spectral position of the peak, highlighted by the grey lines, as well as its linewidth follow similar trends upon increasing the Ni sector. Differences in the magnitude, shape and width of the rotation peaks are attributed to inhomogeneities in the fabricated samples. 

The counterintuitive dependence of the MO activity on the Ni sector arc length suggests that the Faraday rotation is controlled by resonant field enhancement in the vicinity of the Ni section. In order to quantify this enhancement, we compare the MO response of the metamaterial to that of a continuous Ni film. Since the latter exhibits negligible transmission, a straightforward comparison is not instructive. Instead, we use as a figure of merit (FOM) the scattered electric field component which is normal to the incident polarization, normalized to the incident electric field and to the Ni filling factor. The FOM is calculated as $FOM=\frac{|E_{sc}|sin(\phi)}{|E_{inc}|f}\simeq \frac{|E_{sc}|\phi}{|E_{inc}|f}$ (in the small angle approximation), where $\phi$ is the Faraday rotation angle, $f$ is the Ni filling factor, $E_{sc}$ and $E_{inc}$ are the amplitudes of the scattered and incident electric fields, respectively. This FOM allows for a comparison between the MO response of the metamaterial in transmission and that of a continuous Nickel film in reflection. Taking into account that the metamaterial sample with a $90^{\circ}$ Ni sector has a peak rotation of $\sim 0.41$ mrad at $880\,$nm, transmittance of $\sim 10\%$  and a Ni filling factor of $6\%$, the resulting value is $FOM_{MM}\approx 2.16\ 10^{-3}$. In comparison, a continuous, $60\,$ nm thick, Ni film exhibits a reflectance of $70\%$ and a MO rotation of $0.4\,$mrad at the same wavelength resulting in $FOM_{Ni}\approx 0.33\ 10^{-3}$. Hence, the metamaterial leads to an enhancement of magneto-optical activity of $\frac{FOM_{MM}}{FOM_{Ni}}\simeq 6.5$.

\begin{figure}[ht]
	\includegraphics[width=0.5\textwidth]{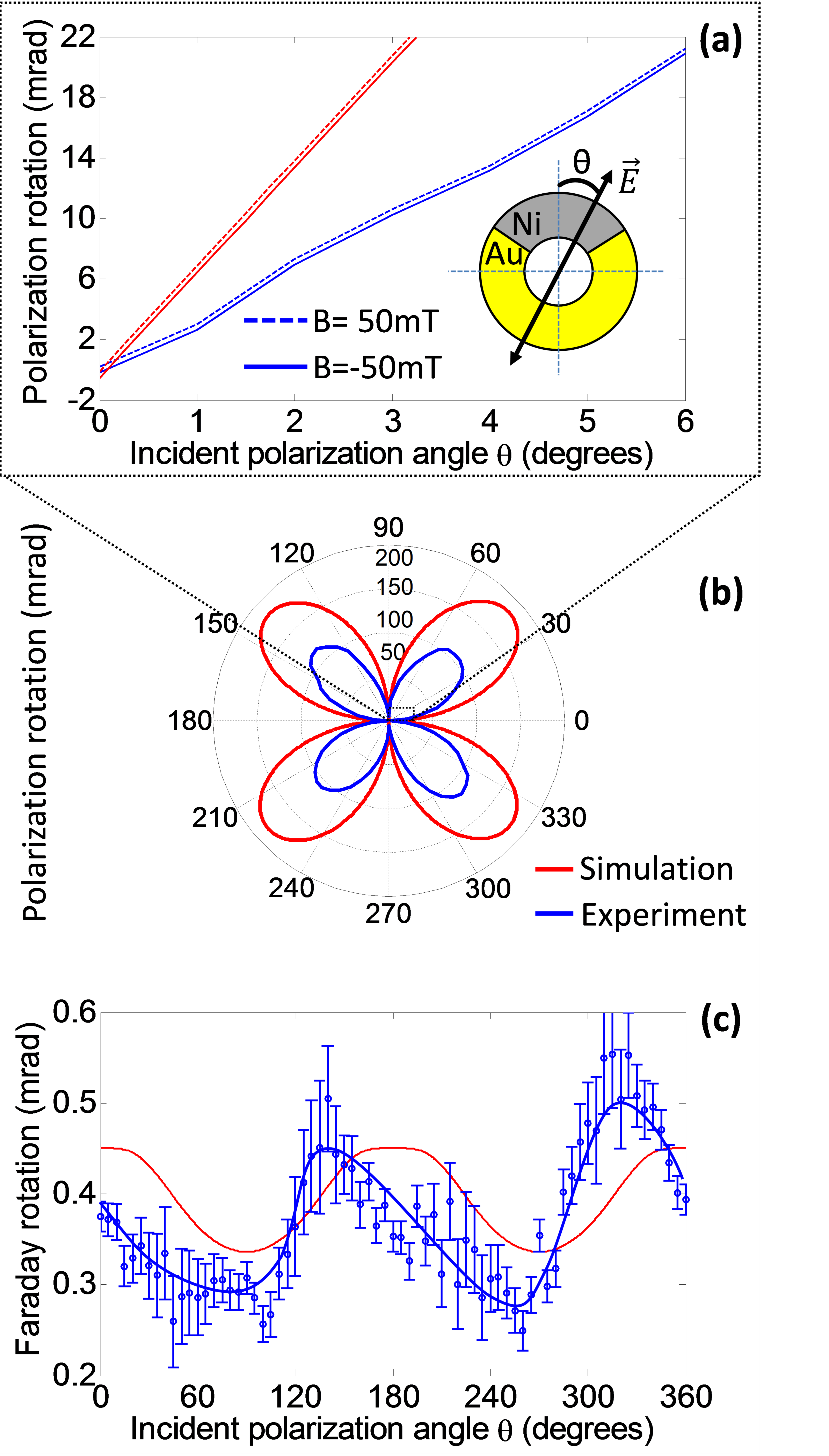}
	\caption{{\bf Sample Anisotropy.} {\bf (a)} {\bf (b)} Rotation of the polarization azimuth of a beam transmitted through the sample ($135^{\circ}$ Ni sector). Red lines represent computational analysis and blue lines stand for experimental results. Solid and dashed lines represent opposite signs of external magnetic fields. $\theta$ is the angle of polarization state in reference to the symmetry axis of the metamaterial array. {\bf (c)} The effect due to the presence of external magnetic field expressed as the difference between solid and dashed lines plotted above. The blue line is a guide for the eye.}
	\label{Fig4}
\end{figure}

% Incident polarization discussion
Owing to the anisotropic nature of the bimetallic rings, the MO response depends strongly on the polarization of the incident wave. The geometrical position and span of the two metal segments create a system with a single plane of symmetry, deviating from which affects both its plasmonic and MO response. In Fig.\ \ref{Fig4}b we can see that the geometrical anisotropy of the sample leads to polarization conversion (in the absence of external static magnetic fields). Varying the angle of polarization azimuth of the excitation beam from $0^{\circ}$ to $360^{\circ}$ results in a four-lobe rosette with the points of zero conversion at $0^{\circ}$, $90^{\circ}$, $180^{\circ}$ and $270^{\circ}$. These points correspond to the polarization eigenstates of the system. In contrast, the effect of anisotropy becomes maximum at n$\times 45^{\circ}$ (for n=1,3,5,7) and it results in a rotation of the incident polarization state by about $100\,$mrad. The introduction of an external magnetic field, creates an additional contribution to the polarization rotation due to magnetically-induced anisotropy, which manifests as an offset of the total rotation. For opposite sign magnetic fields, this effect appears as a splitting of the rotation curve, where the separation between the two curves corresponds to the Faraday effect (see Fig.\ \ref{Fig4}a). The angle of the incident wave's polarization azimuth shifts the current distribution along the ring (between areas of different dielectric properties), which significantly affects the plasmonic enhancement of Faraday rotation. When excitation occurs at angles that drive currents through the more lossy areas (Nickel), the plasmonic resonance is damped and thus, Faraday rotation is reduced by almost $50\%$ (Fig.\ \ref{Fig4}c).  

\begin{figure}[h!]
	\includegraphics[width=0.5\textwidth]{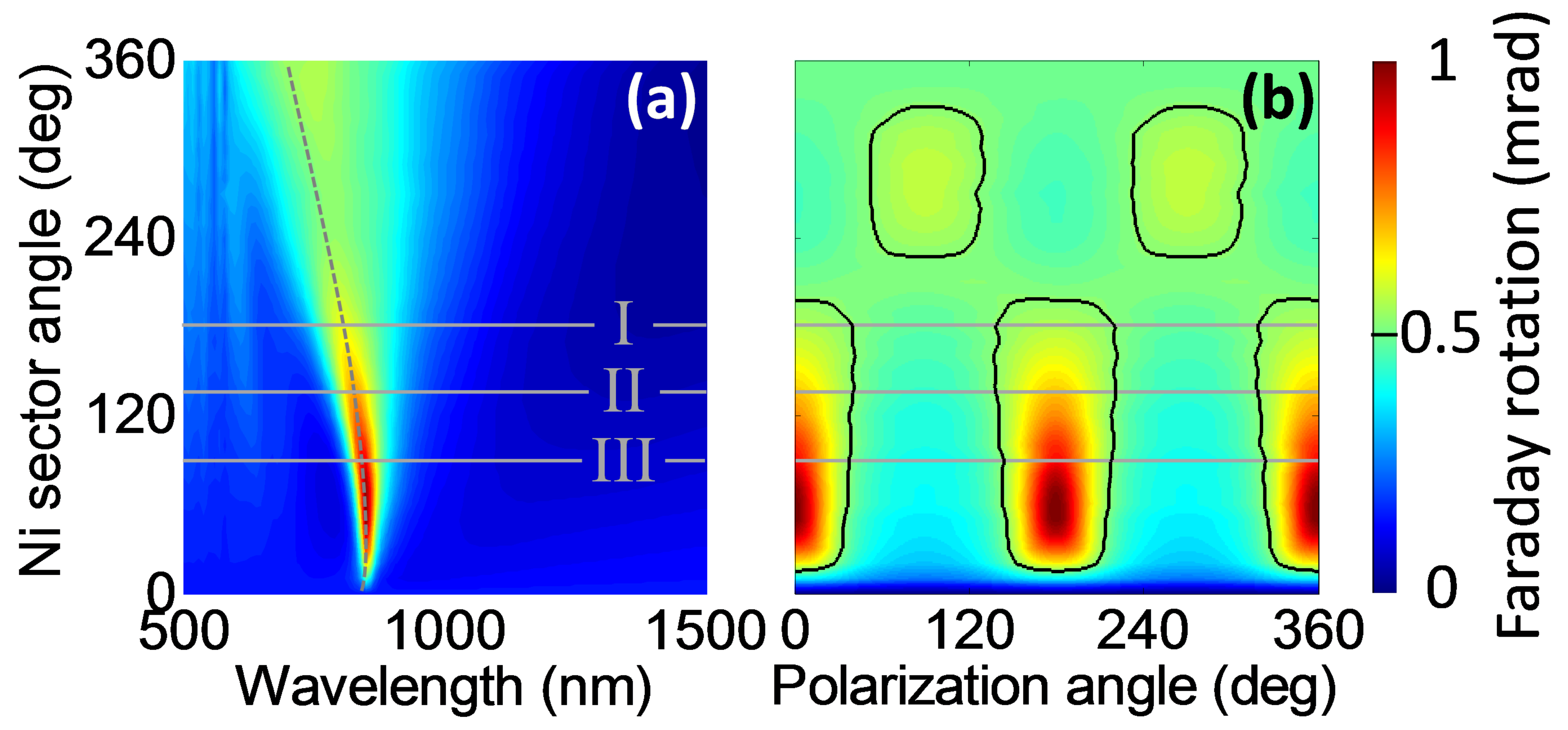}
	\caption{{\bf Bimetallic ring composition properties.} {\bf (a)} Faraday rotation spectra at different ring compositions. {\bf (b)} At resonance, denoted with dashed line on (a), Faraday rotation at different polarization angles of the excitation beam and an external static magnetic field of $100\,$mT.}
	\label{Fig5}
\end{figure}

A detailed numerical study of the effects of ring composition and incidence wave polarization on the Faraday rotation is presented in Figs.\ \ref{Fig5}a \& \ref{Fig5}b, respectively. In accordance to our experimental measurements (see Fig.\ \ref{Fig3}), increasing the size of the Ni sector leads to an increase both in the dissipation loss (hence a decrease in plasmonic field enhancement) and in the filling factor of the MO active material. These two competing mechanisms result in an optimum ring composition with a Ni sector of $\sim 50^{\circ}$, away from which Faraday rotation rapidly decreases (see Fig.\ \ref{Fig5}a). Moreover, rings with relatively small Ni sectors (smaller than $180^{\circ}$) exhibit a strong dependence in the polarization of the incident wave (see Fig.\ \ref{Fig5}b). For polarization along the y-axis (see inset to Fig.\ \ref{Fig2}a), the resonant currents flow mainly in the Au sector, which leads to a strong plasmonic resonance and (through field enhancement) to high values of Faraday rotation. When the polarization is rotated by $90^{\circ}$, the current density in Ni increases, which damps the MO response. On the contrary, for rings with large Ni sectors, the situation reverses and polarization angles normal to the symmetry plane of the ring offer a stronger MO response. In this case, due to the weaker plasmonic resonances, the modulation depth is reduced.

The dependence of the metamaterial MO response on the strength of the external magnetic field was investigated by hysteresis measurements, where the magnetic field intensity is varied, while its direction is  maintained. The measurements reveal an almost linear magnetic field dependence and zero coercivity (Fig.\ \ref{Fig6}). These observations are explained by considering that the magnetic field is orientated normal to the plane of the nano-rings and that the field strength used in the experiment is not sufficient to saturate their magnetization. This behaviour is strongly supported by the micromagnetic simulations (see Methods for details), represented by the solid lines in Fig.\ \ref{Fig6}. The (numerically calculated) magnetization is connected by a linear relation to the (experimentally measured) Faraday rotation allowing, thus, for a direct comparison.    
% ----- Conclusions -----
\section{Conclusion}
We have shown that bimetallic nano-ring resonators can serve as compact building blocks for magneto-optical devices. In particular, we have demonstrated experimentally strong enhancement of the Faraday rotation in metamaterial arrays of such resonators. We also studied the dependence of the effect on external static magnetic fields, polarization of the incident wave, and wavelength, as means of controlling the magneto-optical response of the metamaterial. We expect bimetallic metamaterials to find applications in refractive index and magnetic field sensing, as well as in compact active components for integrated nanophotonic circuits.

% ---------- Methods ----------
\section{Methods}
% fabrication
\textbf{Fabrication.} Samples of different Au-Ni ratios were prepared by a two-step electron-beam lithography method on a glass substrate. Each sample is a regular two-dimensional array with a footprint of $100\ \times 100\  \mu$m$^2$. We have manufactured arrays of bimetallic rings with a period of $400\,$nm and Ni sectors of $90^{\circ}$, $135^{\circ}$ and $180^{\circ}$. The fabrication process is separated in two steps, one for each of the two metals. In each step, metals are thermally evaporated on the substrate with a thickness of $60\,$nm and subsequently lifted-off to reveal the design. A $50\,$nm thick layer of co-polymer deposited underneath 200 nm of PMMA helps increasing the speed of the lift-off process. The Ni part is deposited after Au to minimize any oxidization in the junctions between the two metals. Furthermore, careful design and alignment is crucial in order to achieve good contact at the junctions. Discrepancies between the designed and fabricated samples can be attributed to the surface roughness that is identified mainly in the Ni sector and has an RMS value that varies from $5$ to $20\,$nm between individual unit cells. As Ni tends to form large grains, nanostructures suffer from small deformations, especially in the z-direction, due to the constrains that the e-beam lithography mask enforces in the x-y plane. Nevertheless, such deformations do not alter the plasmonic properties in a significant way as the latter mostly depends on the resonator dimensions in the plane perpendicular to light propagation (x-y plane). Furthermore, these artefacts are inconsistent across each sample and thus lead only to a small inhomogeneous broadening of the plasmonic resonances.   

\textbf{Numerical simulations.} The plasmonic and MO properties of the bimetallic metamaterials were simulated by a finite element method analysis with a commercial solver (COMSOL). An infinite 2D array of resonators is assumed to be positioned on a semi-infinite glass substrate. In the presence of external magnetic fields, the permittivity tensor of Ni becomes non-diagonal  ~\cite{Goto1977,Argyres1955,Zharov2007}. Limiting the magnetic field vector to a direction normal to the plane of the sample reduces the off-diagonal elements to only a single non-zero pair:
\begin{equation}
\epsilon = \epsilon_r\left( \begin{array}{ccc}
1 & -iQ & 0\\
iQ & 1 & 0\\
0 & 0 & 1 \end{array} \right)
\end{equation}
where $\epsilon_r$ is the (isotropic) complex relative permittivity of the magneto-optically active material and $Q$ is a frequency and magnetic field dependent magneto-optical parameter, obtained by interpolating values found in the literature ~\cite{Snow,Goto1977}. 

The Faraday effect highly depends on the transmission properties of the medium, since the rotation of the polarization azimuth is expressed as the angle between the transmitted electric field component which is polarized parallel to incident radiation and the one perpendicular to it. Even small differences between the experimentally and numerically obtained transmittance can lead to large discrepancies in the estimation of the Faraday effect. In order to account for such effects, we normalize the amplitude of the calculated transmitted electric fields to the value we measure experimentally. 

\begin{figure}[htb!]
	\centering
	\includegraphics[width=0.5\textwidth]{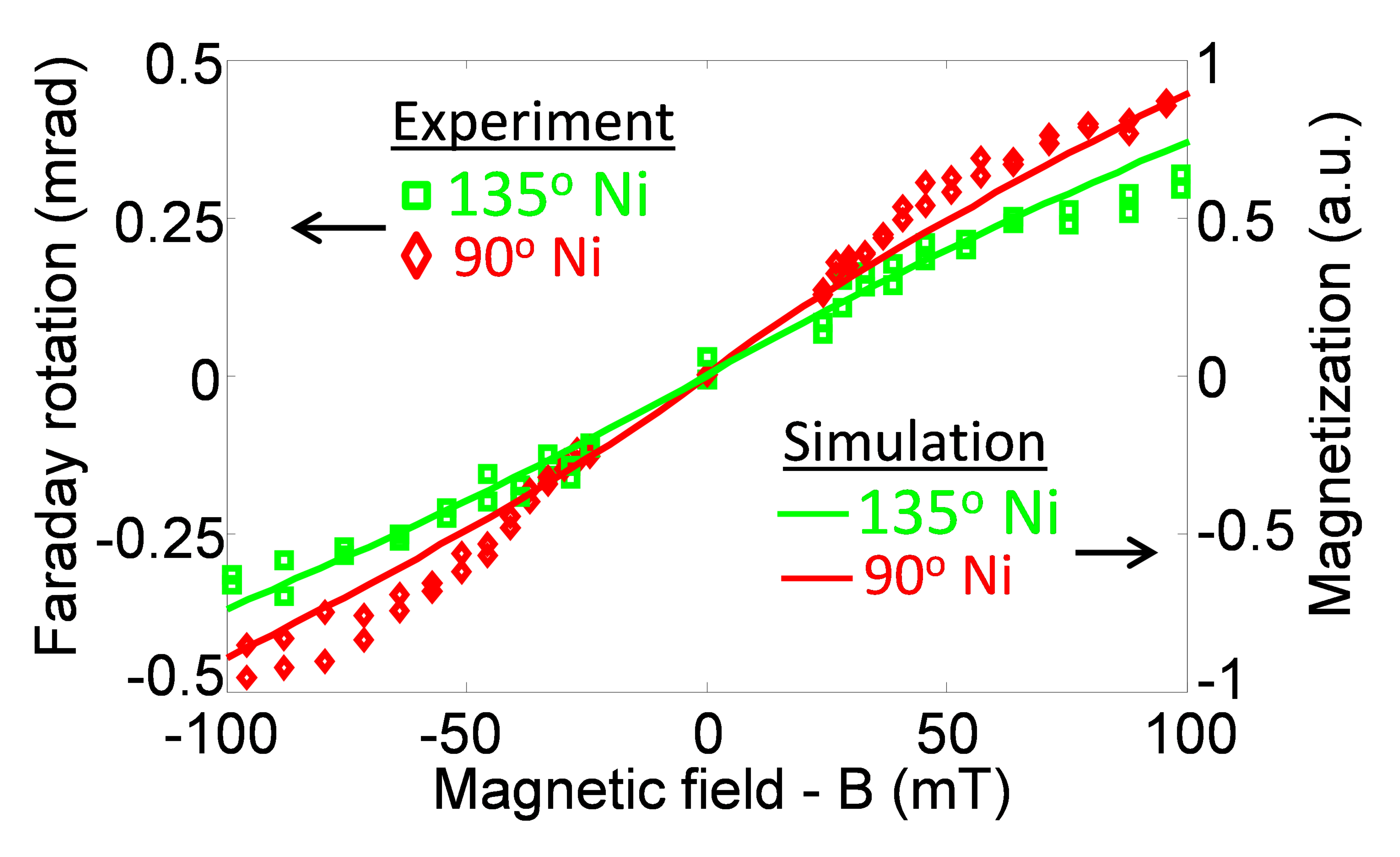}
	\caption[Hysteresis in bimetallic ring resonators]{{\bf Hysteresis in bimetallic ring resonators.} Simulated (solid lines) and experimental measurements (markers) of hysteresis in bimetallic rings with $90^{\circ}$ (red) and $135^{\circ}$ (green) Ni sector. Simulated data correspond to the magnetization on the system while experimental results are presented in the form of Faraday rotation which is proportional to the first.}
	\label{Fig6}
\end{figure}

Hysteresis cycles of bimetallic nano-rings were simulated using OOMMF \cite{OOMMF}. Additional magnetic parameters used are those commonly found in the literature for Ni \cite{Ni_data}, magnetisation saturation is $0.47\,$pJ/m, exchange constant is $8.2\,$pJ/m and a maximum applied external field $H_z$ is $2\times 10^4\,$A/m that is  perpendicular to the plane of the sample. The mesh size was $4\times 4 \times 4\,$nm$^3$, thus allowing accurate simulations while keeping reasonable computational times. The results presented in Fig.\ \ref{Fig6} show a typical hysteresis curve for soft, thick magnetic films (where perpendicular magnetic anisotropy is not present), though modulated due to the non-extended geometry of the Ni sectors.

%Setup
\textbf{Experimental characterization.} The characterization of the linear optical properties for the fabricated samples was performed using a commercial microspectrometer (CRAIC Technologies). A sensitive polarimeter setup, described in \cite{Bennett}, was utilized in order to probe angles of polarization rotation with a precision of $10^{-5}\,$mrads. The setup consists of a polarization modulator (Faraday modulator) positioned between two crossed polarizers. The rotation can be derived by simultaneous detection locked in both the fundamental frequency and second harmonic of the modulator. 
For incident laser power $P_s$, the detected signal can be described by the following equation:
\begin{equation}
	P_{out}=\frac{P_s}{2}(-4A \theta cos(\omega t + \phi) + A^2cos(2\omega t + 2\phi) + A^2),
\end{equation}   
where $A$ is the rotation induced from the Faraday modulator, $\omega$ is its fundamental frequency, $\phi$ is the arbitrary phase of the modulation and $\theta$ is the polarization angle rotation we want to probe.

The two frequency components (fundamental frequency $F$ and $2^{nd}$ harmonic $S$ signal) are recorded by lock-in detection. The rotation is derived as:
\begin{equation}
	\theta=-\frac{F}{S}\frac{A}{4}
\end{equation}   
The rotation spectrum was measured by using a tunable laser source (Spectra-Physics MaiTai) and a photo-detector. The wavelength of the source was varied between $700$ and $1000\,$nm, with a step of $10\,$nm. The external magnetic field was provided by ring-shaped Neodymium magnets placed in the propagation direction, in such way as to form a magnetic field with direction normal to the sample plane. In order to accommodate hysteresis measurements in the setup, the distance between the magnets and the sample was varied in controlled steps.

\section{Acknowledgements}
The authors would like to thank Anibal L. Gonzalez Oyarce for his advice on micromagnetic simulations and F. Javier Garcia de Abajo for numerous fruitful discussions. The authors acknowledge the support of the MOE Singapore (grant MOE2011-T3-1-005), the UK’s Engineering and Physical Sciences Research Council (grants EP/G060363/1), and the Leverhulme Trust.

\end{document}